\documentclass[12pt,letterpaper,nohyper]{JHEP3}
\usepackage{amsmath,amssymb,amsfonts,amsxtra,graphics,graphicx}

\newcommand{\be}{\begin{equation}}
\newcommand{\ee}{\end{equation}}
\newcommand{\beq}{\begin{equation}}
\newcommand{\eeq}{\end{equation}}
\newcommand{\ba}{\begin{array}}
\newcommand{\ea}{\end{array}}
\newcommand{\bea}{\begin{eqnarray}}
\newcommand{\eea}{\end{eqnarray}}
\newcommand{\ben}{\begin{enumerate}}
\newcommand{\een}{\end{enumerate}}
\newcommand{\bean}{\begin{eqnarray*}}
\newcommand{\eean}{\end{eqnarray*}}
\newcommand{\eref}[1]{(\ref{#1})}
\newcommand{\nn}{\nonumber}
\newcommand{\comment}[1]{}

\newcommand{\dd}{{\rm d}}
\newcommand{\AdS}[1]{{\rm AdS}_{#1}}
\newcommand{\GN}{G_4}
\newcommand{\MADM}{M_{\rm ADM}}
\newcommand{\BR}{{\mathbb R}}
\newcommand{\BZ}{{\mathbb Z}}

\title{Cardy \& Kerr}

\author{Vishnu Jejjala \\
Institut des Hautes \'Etudes Scientifiques \\
Le Bois-Marie \\
35, Route de Chartres, 91440 Bures-sur-Yvette, France \\
E-mail: {\tt vishnu@ihes.fr}}

\author{Suresh Nampuri \\
Arnold Sommerfeld Centre for Theoretical Physics \\
Ludwig-Maximilians-Universit\"at M\"unchen \\
Department f\"ur Physik \\
Theresienstr.\ 37, 80333 M\"unchen, Germany \\
E-mail: {\tt Suresh.Nampuri@physik.uni-muenchen.de}}

\abstract{
The Kerr/CFT correspondence employs the Cardy formula to compute the entropy of the left moving CFT states.
This computation, which correctly reproduces the Bekenstein--Hawking entropy of the four-dimensional extremal Kerr black hole, is performed in a regime where the temperature is of order unity rather than in a high-temperature regime.
We show that the comparison of the entropy of the extreme Kerr black hole and the entropy in the CFT can be understood within the Cardy regime by considering a D$0$-D$6$ system with the same entropic properties.
}

\begin{document}
\pagestyle{plain}
\setcounter{page}{1}
\baselineskip16pt

\section{Introduction}
\label{sec:intro}

Black holes provide ideal theoretical laboratories for studying quantum gravity.
The laws of black hole mechanics are formulated in terms of thermodynamics~\cite{bh1}.
Entropy is a thermodynamic quantity that enumerates the quantum states of a black hole.
These quantum states are the true microscopic degrees of freedom that arrest the dynamics of the black hole.
Unfortunately, the underlying theory of gravitational statistical mechanics that underpins black hole thermodynamics remains elusive.
One of the first indications that such a theory might exist is due to Brown and Henneaux~\cite{bh2}, who realised that the asymptotic symmetries of $(2+1)$-dimensional gravity with negative cosmological constant yields a Virasoro algebra with non-zero central extension.
Thus black holes with a near-horizon $\AdS{3}$ factor are described by states in a two-dimensional CFT, a result which anticipates the holographic principle~\cite{holo}.

String theory has sharpened our understanding of black holes considerably and offered profound insights on the microscopic origin of the entropy.
Using duality, enumerating supersymmetric vacua in a weakly coupled description in terms of strings and D-branes correctly reproduces the gravitational entropy in the semiclassical picture of the black hole~\cite{sen,sv}.
However, this description fails to isolate the microstates within the strongly coupled regime:
because a black hole has no hair, the geometry is completely specified by global charges measured at infinity.
In certain highly symmetric settings, the AdS/CFT correspondence~\cite{ads} enables the identification of microstates in gravity and begins to address such long-standing problems as the information paradox~\cite{fuzzball}.
Coarse-graining, or the averaging over an ensemble of states in CFT with appropriate charges, may provide a general mechanism for the origin of gravitational thermodynamics~\cite{babel}.

This research, while of tremendous importance, focuses on black holes that we do not see in Nature.
A first principles accounting of the entropy of the four-dimensional Schwarzschild, Kerr, Reissner--Nordstr\"om, and Kerr--Newman solutions of Einstein gravity in asymptotically flat spacetimes in terms of a microscopic theory is notably lacking.

Recently, it has been argued that by considering a near-horizon limit of the extremal Kerr black hole~\cite{z,bh3}, the entropy can be reproduced from a computation of the central charge in two dimensions~\cite{ghss}.
For the extreme Kerr solution, the angular momentum is specified by the ADM mass:
$J = \GN \MADM^2$.
At fixed polar angle, a slice of the near-horizon extremal Kerr (NHEK) geometry contains a quotient of a warped $\AdS{3}$.
According to~\cite{ghss}, the asymptotic symmetric group contains a left moving Virasoro algebra with central extension $c_L = \frac{12 J}{\hbar}$.
The left moving temperature of the CFT is $T_L = \frac{1}{2\pi}$.
Computing the statistical entropy using the Cardy formula~\cite{c}, one reproduces the gravitational entropy,
\be
S_{\rm BH} = \frac{A}{4\GN\hbar} = \frac{2\pi J}{\hbar} ~,
\label{eq:gravent}
\ee
from counting microscopic degrees of freedom in the CFT:
\be
S_{\rm CFT} = \frac{\pi^2}{3} c_L T_L = \frac{2\pi J}{\hbar} ~.
\label{eq:cardy}
\ee

That these two calculations agree is surprising.
The Cardy formula gives the leading term in the high-temperature expansion of the two-dimensional CFT's partition function and hence properly applies in the regime where the dimensionless temperature is much large than one.
Here, the situation is reversed:
the temperature of the CFT is of order one, and hence the Cardy formula cannot be reliably used to compute the entropy of this system.

A very similar situation vis-\`a-vis the applicability of the Cardy formula occurs in another system in string theory:
the non-supersymmetric extremal D$0$-D$6$ system, which is a bound state of anti-D$0$-branes and D$6$-branes.
When raised to five dimensions, at all points in moduli space where it has a good supergravity description, has a near-horizon geometry whose non-compact part is $\AdS{2}$.
The Cardy formula is a high-temperature expansion formula of the ${\rm CFT}_2$ and hence cannot be used for the ${\rm CFT}_1$ dual to the near-horizon geometry.
The strategy to compute entropy in this case is to use the four-dimensional U-duality of type IIA string theory to go to a point in both charge and moduli space where a good supergravity description of the near-horizon geometry as an $\AdS{3}$ exists wherein the charges are in the correct proportions to satisfy the Cardy limit criterion~\cite{Nampuri:2007gw}.

In fact, for CFTs with gravity duals, applying the Cardy formula in the microscopic theory generically reproduces the gravitational entropy well outside what we would expect to be the formula's regime of validity.
If we consider $n_1$ D$1$-branes and $n_5$ D$5$-branes wrapping a compact manifold ${\cal M} = T^4\ {\rm or}\ {\rm K}3$, the CFT is a sigma model whose target space is a deformation of the orbifold ${\cal M}^N/S_N$, with $c = 6N = 6 n_1 n_5$~\cite{sw}.
In a large-$N$ limit, this is a situation where $c \gg 1$.
This system is dual to an effective fundamental string carrying $n_1$ units of winding and $n_5$ units of momentum.
The oscillator level $N_L = n_1 n_5$ is partitioned among eight bosonic and eight fermionic oscillators, while $N_R = 0$.
The central charge here is $c = 12$.
We are in a regime where the eigenvalue of $L_0$ is much larger than $c$.
This allows for the CFT temperature $T_L \approx \sqrt{L_0}/c \gg 1$.
Use of the Cardy formula is then appropriate and reproduces the entropy of the two-charge black hole~\cite{lm,fuzzball}.
The existence of various winding sectors in the effective string description is crucial to the argument.\footnote{
See also footnote 12 of~\cite{ghss}.}

The goal of this paper is to justify the use of the Cardy formula to compute the entropy of the four-dimensional Kerr black hole.
The note is organised as follows.
In Section~\ref{sec:review}, we very briefly review the proposed Kerr/CFT correspondence.
In Section~\ref{sec:m}, we consider M-theory lifts of neutral black holes in four dimensions and argue that there is a duality frame in which the Cardy formula is valid.
In Section~\ref{sec:modmap}, we map the moduli between the D$0$-D$6$ system and a system of D$0$-D$2$-D$2$-D$4$-D$4$-D$4$-branes.
In Section~\ref{sec:dyn}, we connect the physics of a D$0$-D$6$ system (a bound state of anti-D$0$-branes and D$6$-branes) to known features of the NHEK geometry.
Section~\ref{sec:conc} offers concluding comments.
Technical remarks are collected in the Appendix.

\section{The Kerr/CFT Correspondence}
\label{sec:review}

The Kerr black hole is a solution to the flat space Einstein equations with mass and angular momentum~\cite{kerr}.
The near-horizon extremal limit of the Kerr solution (NHEK), after fixing the polar angle, is the metric of a quotient of warped $\AdS{3}$~\cite{warped}, in analogy to the quotient of pure $\AdS{3}$ yielding the BTZ black hole~\cite{btz}.
The throat geometry enjoys an $SL(2,\BR)$ isometry corresponding to the $\AdS{2}$, and also there is a $U(1)$ corresponding to rotations in the azimuthal angle~\cite{bh3}.

Applying the analysis of Brown and Henneaux~\cite{bh2} to the case of warped $\AdS{3}$,~\cite{ghss} examines the asymptotic symmetry group of the spacetime and finds a chiral Virasoro algebra with central charge $c_L = 12 J$, and a left moving temperature $T_L=\frac{1}{2\pi}$ corresponding to the Frolov--Thorne vacuum~\cite{ft}.
Right moving excitations correspond to deviations from extremality~\cite{ghss,cl}, and as we are considering extremal solutions only $T_R = 0$.

As indicated in~\eref{eq:cardy}, a na\"{\i}ve application of the Cardy formula yields an entropy $S_{\rm CFT} = 2\pi J$~\cite{ghss}.
This is the Kerr/CFT correspondence.
These ideas have been extended to extremal Kerr--Newman--AdS--dS black holes~\cite{hmns}.

In~\cite{ahmrdrs}, it is shown that any solutions to vacuum general relativity that are asymptotically the NHEK metric are as well diffeomorphic to the NHEK metric.
The entropy $S_{\rm CFT}$ measures the degeneracy of the ground state.
There are no excitations above this ground state;
backreaction kills any such modes.
The CFT that is the putative dual of the extremal Kerr black hole is unusual.
There are no dynamics whatsoever.
Note that because the central charge is fixed by the angular momentum, each extremal Kerr black hole corresponds to an unique CFT.
The extremal Kerr solution should be regarded as the vacuum state of the CFT.
This CFT is a chiral CFT and there are indications in~\cite{bdss} that it arises in the DLCQ limit of a two-dimensional non-chiral CFT with the same central charge.
Under the identification that~\cite{bdss} gives for the central charge and the temperature of the thermal state dual to the extreme Kerr black hole, it is found, in agreement with~\cite{ghss}, that $c_L = 12 J$, and the temperature is of order unity.

Let us recapitulate.
The Kerr/CFT correspondence operates in a regime where the Cardy formula is not dependable.
And still it works.
This is a puzzle that we attempt to resolve in the remainder of this paper.

\section{An M-theory Construction}
\label{sec:m}

The aim of this section is to attempt an embedding of the Kerr black hole in type IIA string theory.
String theories offer a description of charged black holes.
These objects have a lower bound on their mass set by the charges that they carry.
When the bound is saturated, the black holes become extremal and no longer radiate so that their temperature is zero, but the solutions may still possess non-zero entropy.
The near-horizon geometry of extremal black holes is an AdS geometry, and so using the AdS/CFT correspondence, these can be regarded as thermal states in the CFT~\cite{witten}.
The global charges encode information about the central charge of the CFT and the excitation above the vacuum (representing pure AdS), which corresponds to the black hole.
Hence, a true microscopic interpretation of a neutral Kerr black hole in string theory would involve identifying a duality frame where it can be seen as a bound state of branes which gravitate to form a black hole with a near-horizon geometry that is dual to a known CFT.

To start, we outline the method~\cite{emparan,Horowitz:2007xq} of connecting a four-dimensional rotating dyonic black hole to a five-dimensional Myers--Perry~\cite{mp} rotating black hole via a M-theory lift.
We will then, following~\cite{emparan}, use a symmetry of this solution to interchange the electric charge and four-dimensional angular momentum and so demonstrate that a static black hole in four dimensions can be lifted via M-theory to a five-dimensional rotating black hole.
By the interchange symmetry this can be brought down again to a four-dimensional neutral rotating black hole.
We first look at the M-theory lifted metric of an extremal dyonically charged four-dimensional rotating black hole.
The metric is of the form
\bea
\label{eq:emet}
&& \dd s^2 = \frac{H_q}{H_p}(dy+{\bf A})^2-\frac{\Delta_\theta}{H_q}(dt+{\bf B})^2+H_p\left(\frac{dr^2}{\Delta}+d\theta^2+\frac{\Delta}{\Delta_\theta}\sin^2\theta\, d\phi^2 \right) ~, \\
&& H_p = r^2 + r\, p + \frac{p^2\, q}{2 (p+q)} + \frac{q\, p^2 \cos\theta}{2 (p+q)}\, \frac{\alpha}{m} ~, \qquad
H_q = r^2 + r\, q + \frac{q^2\, p}{2 (p+q)} - \frac{p\, q^2 \cos\theta}{2 (p+q)}\, \frac{\alpha}{m} ~. \nn
\eea
For brevity, we omit the expressions for $\Delta$, $\Delta_\theta$, ${\bf A}$, and ${\bf B}$;
all the functions that define~\eref{eq:emet} are given explicitly in the Appendix of~\cite{emparan}.
For our purposes it suffices to note that the four-dimensional charges and rotation are as follows:
\be
2 G_4 M = \frac{p+q}{2} ~, \qquad
G_4 J = \frac{(p q)^{\frac32}}{4(p+q)} \frac{\alpha}{m} ~, \qquad
Q^2=\frac{q^3}{4(p+q)} ~, \qquad
P^2= \frac{p^3}{4(p+q)} ~.
\ee
Here, $Q$ and $P$ are the D$0$-brane and D$6$-brane charges, respectively.
In an M-theory lift, the D$0$-brane charge becomes the momentum mode on the M-theory circle, which is transverse to the M-theory monopole to which the D$6$-brane is lifted.
(A discussion of some of the subtleties that should be kept in mind during the M-theory lift may be found in the Appendix.)
The supergravity realisation of this is that of a Taub-NUT/ALE space produced by one or more coincident monopoles with momentum flowing along the Taub-NUT or ALE circle.
These quantities are related to the integral charges as
\be
Q = \frac{2 G_4 N_0}{R} ~, \qquad
P = \frac{R N_6}{4} ~.
\ee
The five-dimensional supergravity picture is a good approximation when the radius $R$ of the Taub-NUT circle is large.
Thus, we will take the limit of $p\to \infty$ with $r$, $m$, $\alpha$, $q\to 0$ and $y\to \infty$ with $pr$, $pm$, $p\alpha$, $pq$, and $\frac{y}{p}$ finite.
The finite parameters may be written as
\bea
p q &=& \frac{\mu}{4} ~, \\
p\alpha &=& \frac{1}{8}(\mu - (a+b)^2)^{\frac{1}{2}}(a-b) ~, \\
p m &=& \frac{1}{8}[\mu (\mu - (a+b)^2)]^{\frac{1}{2}} ~, \\
p r &=& \frac{1}{4}[\rho^2 - \frac{1}{2}(\mu - a^2-b^2 - 8 p m)] ~.
\eea
We write $\psi = \frac{y}{p}$, leading to the identification $\psi \equiv \psi + \frac{4\pi}{N_6}$.
We then recombine the Euler angles into the combinations $\frac{\psi+\phi}{2}$, $\frac{\psi-\phi}{2}$, and $\frac{\theta}{2}$, which we will use as our new Euler angles that we label $\psi$, $\phi$, and $\theta$, respectively.
The periodic identifications in this case are
$(\psi, \phi)\equiv (\psi+\frac{2\pi}{N_6}$,
$\phi + \frac{2\pi}{N_{6}})\equiv (\psi, \phi + 2 \pi)$.
The metric in the limits we have taken reduces to
\bea
\dd s^2 &=& -dt^2 + \frac{\mu}{\Sigma} (dt - a\, \sin^2\theta\, d\psi - b\, \cos^2\theta\, d\phi)^2 + \Sigma \left( \frac{d\rho^2}{\Delta} + d\theta^2 \right) \nn \\
&& + (\rho^2+a^2)\sin^2\theta\, d\psi^2 +(\rho^2+b^2) \cos^2\theta\, d\phi^2 ~,
\label{eq:tn}
\eea
with
\be
\Sigma = \rho^2 + a^2 \cos^2\theta + b^2 \sin^2\theta ~, \qquad
\Delta = \frac{(\rho^2+a^2)(\rho^2+b^2) -\mu \rho^2}{\rho^2} ~.
\ee

The metric~\eref{eq:tn} is a five-dimensional Myers--Perry black hole placed at the tip of the cigar-shaped Taub-NUT.
At large values of the Taub-NUT circle, for $N_6 = 1$ this is a neutral rotating black hole in five-dimensional flat space.
For $N_6 > 1$, it resides at the conifold tip of $\BR^4/\BZ_{N_6}$.
Here $a$ and $b$ are angular momenta rotation parameters corresponding to the two Cartan generators of the $SO(4)$ rotation group in five dimensions given as
\be
J_1 = \frac{\pi\,\mu\,a}{4 G_5 N_6} ~, \qquad
J_2 = \frac{\pi\,\mu\,b}{4 G_5 N_6} ~.
\ee
In terms of the four-dimensional angular momentum $J$ and the charges, these Cartans are given as $J_{1,2} = \frac{N_0 N_6}{2} \pm J$.
The five-dimensional metric enjoys the $\phi \leftrightarrow -\phi$ and $b \leftrightarrow -b$ interchange symmetries.
Under this inversion, $J_2$ is mapped to $-J_2$, and so from the above definitions of the five-dimensional angular momenta, the two Cartan generators are flipped.
Starting from a static four-dimensional D$0$-D$6$ black hole, by flipping, we can end up with a five-dimensional Myers--Perry solution.
In other words, the Myers--Perry black hole in five dimensions is the end point of the lift from four dimensions of a neutral rotating black hole.

The five-dimensional Myers--Perry black hole located at the tip of a Taub-NUT space has the same entropy as the four-dimensional black hole from which it was uplifted.\footnote{
An elaboration of this point and a discussion of the underlying assumptions is in the Appendix.}
Further, because the reflection is a symmetry of the solution, it does not change the horizon area.
This means that the Kerr black hole has the same entropy as the D$0$-D$6$ black hole in another M-theory frame.

Hence, the duality relating the Kerr black hole to a D$0$-D$6$ system is as follows.
If we start with a four-dimensional Kerr black hole with angular momentum $J$, we can effectively think of it as being placed at the tip of a Taub-NUT cigar.
(Here, the horizon of the black hole is assumed to be smaller than the radius of the Taub-NUT circle.)
This is a five-dimensional Myers--Perry black hole with rotation along a Cartan of the rotation group.
It is then reflection dual (for the purposes of entropy counting) to a five-dimensional black hole with rotation axis oriented transverse to the Taub-NUT circle.
The latter is the uplift of a D$0$-D$6$ black hole with one D$6$-brane charge.

Crucially, there is no U-duality frame in which the D$0$-D$6$ black hole exists within the Cardy regime~\cite{Nampuri:2007gw}.
Fortunately, as~\cite{Nampuri:2007gw} demonstrates and as we shall see below, we can justify the Cardy formula by implementing further duality transformations.

One might wonder why the Cardy formula computed for a two-dimensional CFT is applicable at all to a chiral CFT, which is effectively one-dimensional.
To justify this, one must show that the chiral CFT is actually the chiral part of a two-dimensional CFT.
Indications to this effect were noted in~\cite{bdss}, which provided evidence to support the claim that the chiral CFT dual to the Kerr black hole arose in the DLCQ limit of a two-dimensional CFT.
The existence of the two-dimensional CFT allows modular invariance to be a symmetry of the partition function, and the Cardy formula can therefore be derived in a high-temperature limit.
Another way of stating this is to observe that the D$0$-D$6$ system we consider is dual to a D$0$-D$2$-D$2$-D$4$-D$4$-D$4$ non-supersymmetric system, which is a bound state of anti-D$0$-branes, D$2$-branes, and D$4$-branes.
The exact U-duality map between the two configurations is given in~\cite{Nampuri:2007gw}.
The D$0$-D$2$-D$2$-D$4$-D$4$-D$4$ system, at a point in moduli space where the four-dimensional string coupling is strong, can be lifted to five dimensions to obtain a near-horizon geometry that is BTZ in $\AdS{3}$.
This is dual to a two-dimensional CFT, in agreement with~\cite{bdss}.
For the purposes of Bekenstein--Hawking entropy counting, the D$0$-D$6$ and D$0$-D$2$-D$2$-D$4$-D$4$-D$4$ systems are the same.
The U-duality map in~\cite{Nampuri:2007gw} is designed to bring the charges in the final charge configuration of the latter into the Cardy limit.
The map is as follows.

We consider a compactification on $T^4\times T^2$.
Charge vectors are written as:
\be
\vec{Q} = (q_0,-p^1,q_2,q_3,q_4,q_5) ~, \qquad
\vec{P} = (q_1,p^0,p^3,p^2,p^5,p^4) ~.
\ee
Here, $q_0$, $q_1$, $q_2$, $q_3$, $q_4$, and $q_5$ are the D$0$-brane and the D$2$-branes wrapping the two cycles of $T^2$ and $T^4$, respectively, while $p^0$, $p^1$, $p^2$, $p^3$, $p^4$, and $p^5$ are the D$6$-brane and D$4$-branes wrapping $T^4$ and $\Sigma_{2(3)}\times T^2$, where $\Sigma$ represents a two-cycle of $T^4$.
The system consisting purely of D$0$- and D$6$-branes that is our starting point has the charge vectors
\be
\label{charged0d6}
\vec{Q} = (q_0, 0, \ldots, 0) ~, \qquad
\vec{P} = (0, p^0, 0, \ldots, 0) ~.
\ee
The charges satisfy the condition\footnote{
The inner product uses the metric $\eta_{ij} = {\cal H}\oplus {\cal H}\oplus {\cal H}\oplus {\cal H}\oplus {\cal E}_8\oplus {\cal E}_8\oplus {\cal H}\oplus {\cal H}$, where ${\cal H} = \left( \ba{cc} 0 & 1 \cr 1 & 0 \ea \right)$ and ${\cal E}_8$ is the Cartan matrix of $E_8$.}
\be
\label{condad0d6}
|\vec{Q}\cdot \vec{P}| = |q_0p^0|\gg 1 ~.
\ee
An invariant of the duality group is $I = (\vec{Q}\cdot \vec{P})^2 - (\vec{Q}\cdot \vec{Q})(\vec{P}\cdot \vec{P})$.
(The entropy is proportional to $\sqrt{|I|}$ and is therefore also invariant.)
As~\cite{Nampuri:2007gw} explains, the system can be brought into the Cardy regime provided two conditions are satisfied:
\bea
&& p^0 = 0 ~, \label{eq:one} \\
&& I \gg 6 (p^1)^2 (\vec{P}\cdot \vec{P})^2 ~, \label{eq:two}
\eea
where $p^1$ is the charge due to the D$4$-branes wrapping $T^4$.
Under an exact S-duality transformation, one can easily see that the system~\eref{charged0d6} cannot be brought to the Cardy limit.

However, let us consider a nearby set of charges obtained from small changes in the charge vectors that do not affect the leading order entropy.\footnote{
For the change in the charges to be small, we require that
$$
\ba{c}
\left|{\vec{Q}\cdot \Delta \vec{Q} \over \vec{Q}\cdot \vec{P}}\right| \ll 1 ~, \qquad
\left|{\vec{Q}\cdot \Delta \vec{P} \over \vec{Q}\cdot \vec{P}}\right| \ll 1 ~, \qquad
\left|{\vec{P}\cdot \Delta \vec{Q} \over \vec{Q}\cdot \vec{P}}\right| \ll 1 ~, \qquad
\left|{\vec{P}\cdot \Delta \vec{P} \over \vec{Q}\cdot \vec{P}}\right| \ll 1 ~, \cr \cr
\left|{\Delta\vec{Q}\cdot \Delta\vec{Q} \over \vec{Q}\cdot \vec{P}}\right| \ll 1 ~, \qquad
\left|{\Delta\vec{Q}\cdot \Delta\vec{P} \over \vec{Q}\cdot \vec{P}}\right| \ll 1 ~, \qquad
\left|{\Delta\vec{P}\cdot \Delta\vec{P} \over \vec{Q}\cdot \vec{P}}\right| \ll 1 ~.
\ea
$$}
We will show that such a system, which is entropically equivalent to the extremal Kerr black hole, can be taken to the Cardy limit.
Such a small change is
\be
\label{ad0d6}
\vec{Q} = (q_0,0,1,0, \ldots,0) ~, \qquad
\vec{P} = (0,p^0,-1,1, \ldots, 0) ~,
\ee
which corresponds to adding a D$2$-brane, a D$4$-brane, and an anti-D$4$-brane.
In~\eref{ad0d6}, we have activated additional charges lying in the second hyperbolic sublattice.
We could have instead activated the additional charges to lie in any of the other hyperbolic sublattices (or in fact the ${\cal E}_8$ sublattices), and a similar discussion would go through.

Consider the $O(2,2)$ transformation
\be
\label{o2206}
\begin{pmatrix}
1 & 0 & 0 & 0 \cr
0 & 1 & p^0 & 0 \cr
0 & 0 & 1 & 0 \cr
-p^0 & 0 & 0 & 1
\end{pmatrix}
\ee
that acts on the first two $\cal{H}$ sublattices.
This maps the altered charge vectors~\eref{ad0d6} to the form
\be
\label{altd0d6}
\vec{Q}' = (q_0, p^0, 1, -p^0 q_0, 0, \ldots, 0) ~, \qquad
\vec{P}' = (0, 0, -1, 1, 0, \ldots, 0) ~.
\ee
These charge vectors lie within the Cardy limit.
The second entry in $\vec{P}'$ vanishes, and hence~\eref{eq:one} is satisfied.
Also, $p^{1\prime} = p^0$, $(\vec{P}'\cdot \vec{P}') = -2$, so that the condition~\eref{eq:two} is met so long as $|q_0|\gg 1$.

Note that the central charge, $C \simeq |p^{1\prime} (\vec{P}'\cdot \vec{P}')|^2 \simeq (p^0)^2$.
This obeys the condition $C\gg 1$ if $|p^0|\gg 1$.
Alternatively, if $p^0\simeq O(1)$, we can excite additional charges in~\eref{ad0d6} so that, for example, $p^{1\prime} \gg 1$, and thus $C\gg 1$.
Because this system is mapped via string dualities into the Cardy regime, the entropy of the original D$0$-D$6$ system (and therefore the extremal Kerr black hole) can be reliably computed in CFT within a high-temperature expansion.

There is one lacuna in the argument.
Our starting point is a strongly coupled D$0$-D$6$ system.
Under the transformation to a D$0$-D$2$-D$2$-D$4$-D$4$-D$4$ system, we generically end up with a weakly coupled gravity background.
In such a situation, the M-theory description is no longer good.
A U-duality transformation that takes one charge configuration to another also acts on moduli.
We have not carefully examined how the moduli transform under the U-duality.
We will do this in the following section.

\section{Mapping moduli between D$0$-D$6$ and D$0$-D$2$-D$2$-D$4$-D$4$-D$4$}
\label{sec:modmap}

A full analysis of the duality between the D$0$-D$6$ system and the D$0$-D$2$-D$2$-D$4$-D$4$-D$4$ systems involves mapping the moduli between the two systems.
In order to determine the starting point in moduli space for the duality map, we must remember that the D$0$-D$6$ system arose originally from a five-dimensional Kaluza--Klein black hole in Taub-NUT space;
an internal symmetry transformation of the near-horizon geometry accompanied by a downlift to four dimensions gave us the warped $\AdS{3}$ NHEK metric.
The five-dimensional Kaluza--Klein black hole was initially embedded in eleven-dimensional M-theory and then reduced to ten-dimensional type IIA string theory.
In M-theory we have $\BR^{1,3}\times K3\times S_M\times S_1\times S_2$.
In the dual type IIA theory this becomes $\BR^{1,3}\times K3\times S_1\times S_2$.
This in turn, is dual to heterotic string theory on $\BR^{1,3}\times T^3\times S_M\times S_1\times S_2$.
A concise and clear review of the dualities between the heterotic and type IIA theories can be found in~\cite{Dabholkar:2005dt}.
We now use the various dualities between M-theory, type IIA, and heterotic string theory to map all moduli and charges into the heterotic frame where we perform our duality operations.

A few notational comments are useful at this stage.
The $S_M$ circle is the eleven-dimensional M-theory circle with radius $R_{11}$;
the $S_1$ and $S_2$ are circles with radii $R_1$ and $R_2$, respectively.
We can think of these two circles as being $T^2$ at a special point with volume denoted as $V_2$.
Of the $22$ two-cycles of $K3$, $16$ arise as fixed points of $T^4/\BZ_2$ at the orbifold point and are not of interest.
Of the remaining six two-cycles, D$2$-branes in the type IIA frame wrap the cycles corresponding to momentum and winding charges on the $S_1$ circle in the heterotic frame;
these have volumes labelled by $V_{(2)}$ and $V_{(3)}$, respectively.
The volume of the six-dimensional manifold, $K3\times T^2$ is denoted by $V_6$.
The circles themselves may be denoted either by $S$ with an appropriate subscript or by their radii.
The quantities $\ell_{11}$ and $\ell_5$ denote Planck lengths in eleven dimensions and five dimensions, respectively.
The four-dimensional string coupling constant is denoted as $g_4$;
unless explicitly stated, this is represented in the heterotic frame.
All M-theory moduli are measured with respect to the Planck lengths while the string theory moduli are measured with respect to the appropriate string length.

We now employ duality to relate the moduli in the various frames.
By comparing the masses and tensions of the various objects in the different theories, we can first of all relate M-theory moduli to heterotic moduli:
\bea
&& \frac{V_{K3}}{\ell_{11}^4} = (g_4^2)^\frac23 \left( \frac{R_{11}}{\ell_s} \right)^{\frac23} \left( \frac{V_2}{\ell_s^2} \right)^{\frac23} ~, \label{eq:1} \\
&& \frac{R_1}{\ell_{11}} \sqrt\frac{V_{K3}}{\ell_{11}^4} = \frac{R_1}{\ell_s} ~, \label{eq:2} \\
&& \frac{R_2}{\ell_{11}} \sqrt\frac{V_{K3}}{\ell_{11}^4} = \frac{R_2}{\ell_s} ~, \label{eq:3} \\
&& \frac{R_{11}}{\ell_{11}} \sqrt\frac{V_{K3}}{\ell_{11}^4} = \frac{R_{11}}{\ell_s} ~, \label{eq:4} \\
&& \frac{R_{11}}{\ell_5} = \left( \frac{R_{11}}{\ell_s} \right)^{\frac23} \left( \frac{1}{g_4^2} \right)^{\frac13} ~. \label{eq:5}
\eea
In the expressions~\eref{eq:1}--\eref{eq:5}, quantities on the left hand side are written in M-theory while quantities on the right hand side are in heterotic theory.

We now write the relations between type IIA moduli and heterotic moduli:
\bea
&& {R_1\over \ell_s} = \sqrt{R_1\over R_2} {1\over g_4} ~, \label{eq:6} \\
&& {R_2\over \ell_s} = \sqrt{R_2\over R_1} {1\over g_4} ~, \label{eq:7} \\
&& {V_{(2)}\over \ell_s^2} = {R_{11}\over R_1} ~, \label{eq:8} \\
&& {V_{(3)}\over \ell_s^2} = {R_{11} R_{1} \over \ell_s^2} ~, \label{eq:9} \\
&& {V_2\over \ell_s^2} = {1\over g_4^2} ~, \label{eq:10} \\
&& {1\over g_{4,A}^2} = {V_2\over \ell_s^2} ~, \label{eq:11} \\
&& {V_6\over \ell_s^6} = {1\over g_4^2}{R_{11}^2\over \ell_s^2} ~. \label{eq:12}
\eea
In the expressions~\eref{eq:6}--\eref{eq:12}, quantities on the left hand side are in type IIA while quantities on the right hand side are in heterotic theory.
In particular, we caution that $\ell_s$ is either the type IIA or the heterotic string length depending on which side of the equation it appears.

We want gravity to be weak in five dimensions so that supergravity is a good approximation.
This demands that ${R_{11}\over \ell_{11}}$ is large.
From \eref{eq:4}, we observe that for small dimensionless values of the $K3$ volume in M-theory, we require that $R_{11}$ is large in heterotic string units.
From~\eref{eq:10}, we conclude that if the volume of the $T^2$ is large in type IIA units, we have a weakly coupled four-dimensional heterotic theory.
We also see that demanding that the various K\"ahler moduli be large in type IIA units also requires them to be large in heterotic string units and for $R_{11}\gg R_1$.
This defines our initial point in heterotic moduli space as a weakly coupled low-curvature point.

For reference, we write down the general $O(2,2)$ Narain lattice charges in both the heterotic and the type IIA frames as follows:
\be
(Q|P)=(q_0,-p^1,q_2,q_3|q_1,p^0,p^3,p^2)_A = (n,w,\tilde{n},\tilde{w}|W,N,\tilde{W},\tilde{N})_H ~.
\ee
The type IIA charges are electric and magnetic in the heterotic frame and written in IIA language.
The four electric charges correspond to the D$0$-brane, the D$4$-brane wrapping $K3$, and D$2$-branes wrapping two of the $22$ cycles of $K3$.
The magnetic charges correspond to a D$2$-brane wrapping the $T^2$, a D$6$-brane wrapping $K3\times T^2$, and D$4$-branes wrapping the four-cycles dual to the electric D$2$-branes.
In heterotic language, the electric charges are the momentum and the fundamental string winding numbers on $R_{11}$ and the momentum and winding numbers on $R_1$, while the magnetic charges and the NS$5$-brane and Kaluza--Klein monopole dual to $S_M$ and the NS$5$-brane and Kaluza--Klein monopole dual to $S_1$.

The initial charge configuration is $(q_0,0,0,0|0,1,0,0)$.
We first perform the T-duality operation of flipping $S_1$ and $S_2$ so that the second lattice in the charge space now corresponds to $S_2$ with a radius $R_2$, which is large in heterotic string units.
We then perform a $R_2\rightarrow {\ell_s^2\over R_2}$ T-duality operation making the radius small in string units.
We will denote the new radius by the same symbol $R_2$.
After an `infinitesimal' change in the charges, which is sufficiently small to keep the leading order entropy unchanged, we obtain the configuration $(q_0,0,1,0|0,1,-1,1)$.
We then apply the duality map~\eref{o2206} with $p_0=1$ to obtain the charge configuration $(q_0,1,1,-q_0|0,0,-1,1)$.
The duality map operates to give new values of the radii of the $S_M$ and $S_2$ circles:
\bea
R'_{11} &=& \frac{R_{11}}{\sqrt{1+R^2_{11}R_2^2/\ell_s^4}} ~, \\
R'_2 &=& \frac{R_2}{\sqrt{1+R^2_{11}R_2^2/\ell_s^4}} ~.
\eea
The primed quantities here are the new values of the radii in heterotic string units.
If we initially start off with a value of $R_2$ large enough so that after the T-duality radius inversion, its radius is small enough to render the denominator on the right hand side of both of the above expressions to be approximately unity, the duality map effectively leaves the radii unaffected.
We can then perform another T-duality inversion on $S_2$ and exchange $S_1$ and $S_2$ leaving all moduli effectively unchanged.
Since, we started off with moduli corresponding to the radius of the M-theory circle being large in eleven-dimensional Planck units, and therefore by extension in five-dimensional Planck units.\footnote{
We have $\ell_5 = g_4^{\frac23} \ell_{11}$.
Hence, for a weakly coupled heterotic string theory, the five-dimensional Planck length is smaller than the eleven-dimensional Planck length.}.
All K\"ahler moduli of the compact manifold are large in string units, so we end up with a weakly coupled five-dimensional supergravity theory after the duality transformations.

We now focus on the attractor values of the moduli in supergravity, measured in the heterotic frame.
The radii of $S_M$ and $S_1$ are fixed in the attractor background to the same value of ${R_{11}\over \ell_s}={R_1\over \ell_s}=\sqrt{q_0}$.
Hence, one can easily see from the duality equations that provided the constant value of $R_2$ is chosen to be large in heterotic string units and bounded below by $g_4^{-2} q_0^{\frac13}$, the attractor geometry is in a weakly coupled supergravity regime.
We now have a dyonic charge configuration in the Cardy limit at a point in moduli space where it admits a weakly coupled five-dimensional supergravity description.
The near-horizon geometry for this system is a BTZ in $\AdS{3}$, and consequently, one can think of the black hole as a thermal state in the boundary CFT${}_2$ dual to the $\AdS{3}$;
the Cardy formula is applied to compute the Bekenstein--Hawking entropy formally.

\section{Dynamics of Near-horizon D$0$-D$6$ and Extreme Kerr}
\label{sec:dyn}

Here, we discuss the implications of the duality between the D$0$-D$6$ black hole and the extreme Kerr black hole in four dimensions.

The extremal D$0$-D$6$ system when uplifted to five dimensions has a near-horizon geometry of $\AdS{2}\times S^3$~\cite{Nampuri:2007gw}, and consequently, its holographic field theory dual is a ${\rm CFT}_1$.
The extremal D$0$-D$6$ black hole cannot be viewed as an excitation in $\AdS{2}$ unlike black holes with a near-horizon geometry with an $\AdS{3}$ factor.
To see this, note that BTZ black holes in $\AdS{3}$ have a near-horizon geometry of quotiented/thermal $\AdS{3}$, and the temperature of the $\AdS{3}$ is a measure of its excitation above the vacuum, which is pure $\AdS{3}$.
However, a black hole which has a near-horizon geometry of $\AdS{2}$ cannot be viewed as a thermal excitation above pure $\AdS{2}$, and hence is the only solution modulo diffeomorphisms to an $\AdS{2}$ asymptotic geometry of given radius and with the given charges.

For black holes with near-horizon geometry $\AdS{3}$, which upon Kaluza--Klein reduction on $S^1$ have an $\AdS{2}$ factor, there is a gauge field in the two-dimensional geometry arising from the momentum mode along the Kaluza--Klein circle.
This serves to distinguish solutions in two dimensions which are Kaluza--Klein reduced versions of excitations in $\AdS{3}$ over the reduction of the vacuum $\AdS{3}$ solution.\footnote{
See~\cite{Strominger:1998yg,Balasubramanian:2003kq,Gupta:2008ki,Sen:2008yk} for details.}
A near-horizon locally $\AdS{2}$ geometry without a gauge field can be thought of as the lowest state in this spectrum.
Since it is the first state above pure vacuum, which must be globally $\AdS{2}$, it corresponds to a CFT excitation of the order of the central charge itself.
Hence, even if the boundary CFT${}_1$ can be thought of as the DLCQ limit of a two-dimensional CFT, this state does not satisfy the Cardy limit criterion.
Under a subgroup of the U-duality group which changes the number of D$0$-branes and D$6$-branes without exciting the other charges, the near-horizon geometry remains a local $\AdS{2}$ under the M-theory lift.
This means that the Cardy formula will {\em never} be applicable at any point of this duality orbit.
Since the system is dual to an extremal Kerr black hole, and the position of the system within the duality orbit is reflected by the temperature of the thermal state holographically dual to the Kerr black hole, it follows that using the duality subgroup, the temperature of the thermal state can never be brought into the Cardy regime.
Technically speaking, the radius of the $\AdS{2}$ is determined by both charges, and there is no third charge as in the $P$ of the D$1$-D$5$-$P$ system to label excitations in the dual field theory with a central charge given by the AdS radius.

Now, since the temperature of the thermal ensemble dual to the black hole in the CFT is uniquely characterised by the charge quantum numbers of the system, the fundamental string excitations of the D$0$-D$6$ system all correspond to the same macrostate (since they have the same charges as the ground state of the D$0$-D$6$ system).
Hence, they are microstates in the thermal ensemble corresponding to the macrostate specified by the D$0$-brane and D$6$-brane quantum numbers of the black hole.
The extremal Kerr black hole that is dual to the extremal D$0$-D$6$ system and is completely specified by
$J = N_0 N_6$
has a near-horizon geometry of a warped $\AdS{3}$ whose radius is also specified by $J$.
Consequently, the central charge of the holographic CFT is given by $J$ as well, with no other quantum number for labelling excitations above the vacuum.
So one would expect that the extremal Kerr black hole cannot be viewed as an excitation above a vacuum state in the CFT, and its near-horizon geometry, the NHEK, must be unique (up to diffeomorphisms) as a geometry that has the same asymptotics as the NHEK and the same charges.
This logic agrees with that of~\cite{ahmrdrs}.
The fact that the NHEK is a vacuum state of the holographically dual CFT is consistent with the known dynamics of the near-horizon D$0$-D$6$ system.
An exploration of the duality between the Kerr black hole and the D$0$-D$2$-D$2$-D$4$-D$4$-D$4$ system\footnote{
For a thorough discussion and elaboration of this duality, see Section 4 of~\cite{Nampuri:2007gw}.}
can yield information on what the excited states of this CFT are in the Kerr frame.
This is one of the most exciting aspects of this duality.

\section{Summary}
\label{sec:conc}

The near-horizon geometry of a Kerr black hole is a quotient of warped $\AdS{3}$ at fixed values of the polar angle.
From the thermodynamics of black holes in warped $\AdS{3}$~\cite{warped}, there have been indications that this geometry is holographically dual to a two-dimensional CFT with well-specified left and right moving central charges.

The black hole under this holographic correspondence is a thermal state in the CFT with a left moving temperature of order unity and zero right moving temperature~\cite{ghss}.
This places it far beyond the regime of applicability of the Cardy formula.
In~\cite{ahmrdrs}, the near-horizon metric of the NHEK solution was shown to be unique up to diffeomorphisms.
Hence, the NHEK must be regarded as the vacuum of the CFT.

Using~\cite{emparan,Horowitz:2007xq}, we relate the Kerr black hole to a D$0$-D$6$ black hole in four dimensions with a single unit of D$6$-brane charge.
The four-dimensional U-duality symmetry of type IIA is used to transform this system to a D$0$-D$2$-D$2$-D$4$-D$4$-D$4$ configuration with asymptotic $\AdS{3}$ near-horizon geometry with the charges lying in the Cardy regime.
This brings us into a regime in charge space where the Cardy formula allows us to compute the black hole entropy (modulo a caveat about the moduli), and it agrees with the known gravity computation.
This justifies the calculation of the entropy and provides a string theory embedding of the Kerr/CFT correspondence.

\section{Appendix}
\label{sec:App}

Here, we examine more closely the connection between the D$0$-D$6$ system and the Kerr black hole and briefly comment on features of the M-theory lift.
Putting a four-dimensional black hole at the tip of the Taub-NUT involves a number of implicit assumptions.\footnote{
We are grateful to Jan de Boer for encouraging us to expand on this point.}
The D$0$-D$6$ black hole we start with is a configuration in strongly coupled four-dimensional type IIA gravity.
To be precise, the four-dimensional coupling constant is the radius of the five-dimensional Taub-NUT circle.
The volume of the compact six-dimensional space in eleven-dimensional Planck units is unfixed;
the volume in string units is $g_{10}^{-2}$ times the volume in eleven-dimensional Planck units~\cite{dm}.
This means that $g_{10}^2$ times the volume in string units is unfixed.
So $g_4^2$ is also unfixed.
The five-dimensional radius of the Taub-NUT is therefore not frozen by the attractor mechanism and thus cannot affect the entropy.
In principle, one can slide the Taub-NUT radius or the four-dimensional coupling constant from small to large values and go from four-dimensional supergravity to five-dimensional supergravity.
This allows us to lift to M-theory where there is a weakly coupled description in terms of the Myers--Perry black hole.

In the near-horizon geometry of a Myers--Perry black hole, the attractor mechanism fixes all moduli which contribute to the entropy in terms of an invariant constructed out of the charges.
This invariant is the same for all four-dimensional charge configurations in the same U-duality orbit.
As there is only one possible continuous U-duality invariant of the duality group, the large charge entropy computed for the systems must be the same in four and five dimensions (although, in principle, it could have been the case that from the field theory point of view the leading order entropy suffered large corrections as one moved along the Taub-NUT radius scale).
The attractor mechanism, which is a supergravity result, is being implicitly used in our argument for the equivalence of entropy of the two systems.

Two U-dual four-dimensional configurations lifted up to five dimensions will have the same entropy.
Properly speaking, this only indicates a microscopic non-renormalisation of the leading order entropy.
In particular, we do not yet claim that the microstates of the two systems are in one-to-one correspondence.

Furthermore, in the attractor region, the four-dimensional string couping constant or the radius of the M-theory circle is a product of the asymptotic value of the coupling constant and the ratio of the D$0$-brane and D$6$-brane charges.
Therefore it is a flat direction for the entropy since one can tune the asymptotic value of the coupling to change its value in the attractor region.
Consequently, one can tune it to small values to reduce the five-dimensional configuration after reflection to a weakly coupled four-dimensional Kerr solution.

\section*{Acknowledgements}
We thank Atish Dabholkar, Roberto Emparan, Michael Haack, Daniel Jafferis, Chethan Krishnan, Magdalena Larfors, Samir Mathur, Djordje Minic, Noriaki Ogawa, Masaki Shigemori, Joan Sim\'on, and Sandip Trivedi for discussions.
We would especially like to thank Jan de Boer and Monica Guica for their valuable inputs.
This work was completed while the authors were at the seventh Simons Workshop in Stony Brook.
VJ as well expresses gratitude to the LPTHE, Jussieu, the Centro de Ciencias de Benasque Pedro Pascual, and the Physics Department at Virginia Tech for their generous hospitality.
SN specially wishes to thank Sandip Trivedi for initiating questions about this topic and for a most enriching year long discussion.
SN is supported in part by the European Community's human potential program under contract MRTN-CT-2004-005104 ``Constituents, Fundamental Forces and Symmetries of the Universe,'' the Excellence Cluster ``The Origin and the Structure of the Universe'' in Munich, and the German Research Foundation (DFG) within the Emmy-Noether-Program (grant number: HA 3448/3-1).


{\footnotesize
\begin{thebibliography}{99}

\bibitem{bh1}
J.~D.~Bekenstein,
``Black holes and entropy,''
Phys.\ Rev.\ D {\bf 7}, 2333 (1973);
J.~M.~Bardeen, B.~Carter, and S.~W.~Hawking,
``The four laws of black hole mechanics,''
Commun.\ Math.\ Phys.\ {\bf 31}, 161 (1973);
S.~W.~Hawking,
``Particle creation by black holes,''
Commun.\ Math.\ Phys.\ {\bf 43}, 199 (1975)
[Erratum-ibid.\ {\bf 46}, 206 (1976)].

\bibitem{bh2}
J.~D.~Brown and M.~Henneaux,
``Central charges in the canonical realization of asymptotic symmetries: An example from three-dimensional gravity,''
Commun.\ Math.\ Phys.\ {\bf 104}, 207 (1986).

\bibitem{holo}
G.~'t Hooft,
``Dimensional reduction in quantum gravity,''
arXiv:gr-qc/9310026;

L.~Susskind,
``The world as a hologram,''
J.\ Math.\ Phys.\ {\bf 36}, 6377 (1995)
[arXiv:hep-th/9409089].

\bibitem{sen}
A.~Sen,
``Extremal black holes and elementary string states,''
Mod.\ Phys.\ Lett.\ A {\bf 10}, 2081 (1995)
[arXiv:hep-th/9504147].

\bibitem{sv}
A.~Strominger and C.~Vafa,
``Microscopic origin of the Bekenstein-Hawking entropy,''
Phys.\ Lett.\ B {\bf 379}, 99 (1996)
[arXiv:hep-th/9601029].

\bibitem{ads}
J.~M.~Maldacena,
``The large N limit of superconformal field theories and supergravity,''
Adv.\ Theor.\ Math.\ Phys.\ {\bf 2}, 231 (1998)
[Int.\ J.\ Theor.\ Phys.\ {\bf 38}, 1113 (1999)]
[arXiv:hep-th/9711200];
S.~S.~Gubser, I.~R.~Klebanov, and A.~M.~Polyakov,
``Gauge theory correlators from non-critical string theory,''
Phys.\ Lett.\ B {\bf 428}, 105 (1998)
[arXiv:hep-th/9802109];
E.~Witten,
``Anti-de Sitter space and holography,''
Adv.\ Theor.\ Math.\ Phys.\ {\bf 2}, 253 (1998)
[arXiv:hep-th/9802150].

\bibitem{fuzzball}
S.~D.~Mathur,
``The fuzzball proposal for black holes: An elementary review,''
Fortsch.\ Phys.\ {\bf 53}, 793 (2005)
[arXiv:hep-th/0502050].

\bibitem{babel}
V.~Balasubramanian, J.~de Boer, V.~Jejjala, and J.~Simon,
``The library of Babel: On the origin of gravitational thermodynamics,''
JHEP {\bf 0512}, 006 (2005)
[arXiv:hep-th/0508023].
For a review, see
V.~Balasubramanian, J.~de Boer, S.~El-Showk, and I.~Messamah,
``Black holes as effective geometries,''
Class.\ Quant.\ Grav.\ {\bf 25}, 214004 (2008)
[arXiv:0811.0263 [hep-th]].

\bibitem{z}
O.~B.~Zaslavsky,
``Horizon/matter systems near the extreme state,''
Class.\ Quant.\ Grav.\ {\bf 15}, 3251 (1998)
[arXiv:gr-qc/9712007].

\bibitem{bh3}
J.~M.~Bardeen and G.~T.~Horowitz,
``The extreme Kerr throat geometry: A vacuum analog of AdS(2) x S(2),''
Phys.\ Rev.\ D {\bf 60}, 104030 (1999)
[arXiv:hep-th/9905099].

\bibitem{ghss}
M.~Guica, T.~Hartman, W.~Song, and A.~Strominger,
``The Kerr/CFT correspondence,''
arXiv:0809.4266 [hep-th].

\bibitem{c}
G.~H.~Hardy and S.~Ramanujan,
``Asymptotic formulae in combinatory analysis,''
Proc.\ London Math.\ Soc.\ (2) {\bf 17}, 75 (1918);
H.~Rademacher,
``On the partition function p(n),''
Proc.\ London Math.\ Soc.\ {\bf 43}, 241 (1937);
J.~L.~Cardy,
``Operator content of two-dimensional conformally invariant theories,''
Nucl.\ Phys.\ B {\bf 270}, 186 (1986).

\bibitem{Nampuri:2007gw}
S.~Nampuri, P.~K.~Tripathy, and S.~P.~Trivedi,
``Duality symmetry and the Cardy limit,''
JHEP {\bf 0807}, 072 (2008)
[arXiv:0711.4671 [hep-th]].

\bibitem{sw}
J.~de Boer,
``Six-dimensional supergravity on S**3 x AdS(3) and 2d conformal field theory,''
Nucl.\ Phys.\ B {\bf 548}, 139 (1999)
[arXiv:hep-th/9806104];
N.~Seiberg and E.~Witten,
``The D1/D5 system and singular CFT,''
JHEP {\bf 9904}, 017 (1999)
[arXiv:hep-th/9903224];
F.~Larsen and E.~J.~Martinec,
``U(1) charges and moduli in the D1-D5 system,''
JHEP {\bf 9906}, 019 (1999)
[arXiv:hep-th/9905064].

\bibitem{lm}
O.~Lunin and S.~D.~Mathur,
AdS/CFT duality and the black hole information paradox,''
Nucl.\ Phys.\ B {\bf 623}, 342 (2002)
[arXiv:hep-th/0109154].

\bibitem{kerr}
R.~P.~Kerr,
``Gravitational field of a spinning mass as an example of algebraically special metrics,''
Phys.\ Rev.\ Lett.\ {\bf 11}, 237 (1963).
For a review, see
M.~Visser,
``The Kerr spacetime: A brief introduction,''
arXiv:0706.0622 [gr-qc].

\bibitem{warped}
S.~Detournay, D.~Orlando, P.~M.~Petropoulos, and P.~Spindel,
``Three-dimensional black holes from deformed anti de Sitter,''
JHEP {\bf 0507}, 072 (2005)
[arXiv:hep-th/0504231];
I.~Bengtsson and P.~Sandin,
``Anti-de Sitter space, squashed and stretched,''
Class.\ Quant.\ Grav.\ {\bf 23}, 971 (2006)
[arXiv:gr-qc/0509076];
D.~Anninos, W.~Li, M.~Padi, W.~Song, and A.~Strominger,
``Warped AdS3 black holes,''
JHEP {\bf 0903}, 130 (2009)
[arXiv:0807.3040 [hep-th]].

\bibitem{btz}
M.~Banados, C.~Teitelboim, and J.~Zanelli,
``The black hole in three-dimensional space-time,''
Phys.\ Rev.\ Lett.\ {\bf 69}, 1849 (1992)
[arXiv:hep-th/9204099];
M.~Banados, M.~Henneaux, C.~Teitelboim, and J.~Zanelli,
``Geometry of the (2+1) black hole,''
Phys.\ Rev.\ D {\bf 48}, 1506 (1993)
[arXiv:gr-qc/9302012].

\bibitem{ft}
V.~P.~Frolov and K.~S.~Thorne,
``Renormalized stress-energy tensor near the horizon of a slowly evolving, rotating black hole,''
Phys.\ Rev.\ D {\bf 39} (1989) 2125.

\bibitem{cl}
Y.~Matsuo, T.~Tsukioka, and C.~M.~Yoo,
``Another realization of Kerr/CFT correspondence,''
arXiv:0907.0303 [hep-th];
``Yet another realization of Kerr/CFT correspondence,''
arXiv:0907.4272 [hep-th];
A.~Castro and F.~Larsen,
``Near extremal Kerr entropy from AdS2 quantum gravity,''
arXiv:0908.1121 [hep-th].

\bibitem{hmns}
T.~Hartman, K.~Murata, T.~Nishioka, and A.~Strominger,
``CFT duals for extreme black holes,''
JHEP {\bf 0904}, 019 (2009)
[arXiv:0811.4393 [hep-th]].

\bibitem{ahmrdrs}
A.~J.~Amsel, G.~T.~Horowitz, D.~Marolf, and M.~M.~Roberts,
``No dynamics in the extremal Kerr throat,''
arXiv:0906.2376 [hep-th];
O.~J.~C.~Dias, H.~S.~Reall, and J.~E.~Santos,
``Kerr-CFT and gravitational perturbations,''
arXiv:0906.2380 [hep-th].

\bibitem{bdss}
V.~Balasubramanian, J.~de Boer, M.~M.~Sheikh-Jabbari, and J.~Simon,
``What is a chiral 2d CFT? And what does it have to do with extremal black holes?,''
arXiv:0906.3272 [hep-th].

\bibitem{witten}
E.~Witten,
``Anti-de Sitter space, thermal phase transition, and confinement in gauge theories,''
Adv.\ Theor.\ Math.\ Phys.\ {\bf 2}, 505 (1998)
[arXiv:hep-th/9803131].

\bibitem{emparan}
R.~Emparan and A.~Maccarrone,
``Statistical description of rotating Kaluza-Klein black holes,''
Phys.\ Rev.\ D {\bf 75}, 084006 (2007)
[arXiv:hep-th/0701150].

\bibitem{Horowitz:2007xq}
G.~T.~Horowitz and M.~M.~Roberts,
``Counting the microstates of a Kerr black hole,''
Phys.\ Rev.\ Lett.\ {\bf 99} (2007) 221601
[arXiv:0708.1346 [hep-th]].

\bibitem{mp}
R.~C.~Myers and M.~J.~Perry,
``Black holes in higher dimensional space-times,''
Annals Phys.\ {\bf 172}, 304 (1986).

\bibitem{Dabholkar:2005dt}
A.~Dabholkar, F.~Denef, G.~W.~Moore, and B.~Pioline,
``Precision counting of small black holes,''
JHEP {\bf 0510} (2005) 096
[arXiv:hep-th/0507014].

\bibitem{Strominger:1998yg}
A.~Strominger,
``AdS(2) quantum gravity and string theory,''
JHEP {\bf 9901}, 007 (1999)
[arXiv:hep-th/9809027].

\bibitem{Balasubramanian:2003kq}
V.~Balasubramanian, A.~Naqvi, and J.~Simon,
``A multi-boundary AdS orbifold and DLCQ holography: A universal holographic description of extremal black hole horizons,''
JHEP {\bf 0408} (2004) 023
[arXiv:hep-th/0311237].

\bibitem{Gupta:2008ki}
R.~K.~Gupta and A.~Sen,
``Ads(3)/CFT(2) to Ads(2)/CFT(1),''
JHEP {\bf 0904}, 034 (2009)
[arXiv:0806.0053 [hep-th]].

\bibitem{Sen:2008yk}
A.~Sen,
``Entropy function and AdS(2)/CFT(1) correspondence,''
JHEP {\bf 0811}, 075 (2008)
[arXiv:0805.0095 [hep-th]].

\bibitem{dm}
A.~Dhar and G.~Mandal,
``Probing 4-dimensional nonsupersymmetric black holes carrying D0- and D6-brane charges,''
Nucl.\ Phys.\ B {\bf 531}, 256 (1998)
[arXiv:hep-th/9803004].

\end{thebibliography}}
\end{document}